\documentclass{ws-procs9x6}

\def \dndy  {dN/dy} 
     
\def \kpi {K^{-}/\pi^{-}}

\begin{document}

\title{Collision Dynamics at RHIC}

\author{Olga Barannikova}
\address{(for the STAR Collaboration)\\
 Department of Physics, Purdue University, \\ 
525 Northwestern Ave., W. Lafayette, IN 47907, USA\\ 
E-mail: barannik@physics.purdue.edu}

\maketitle

\abstracts{
Measurements of a variety of hadron species in  Au+Au collisions at 200 GeV by the STAR experiment are used to investigate  properties of such collisions.  We present a study of particle yields and  transverse momentum spectra within the framework of chemical and local kinetic equilibrium models. The extracted freeze-out  properties are studied  as function of collision centrality. Those properties together with inferred initial conditions  provide  insights about collision dynamics at RHIC. 
}

\maketitle
\setcounter{page}{1}

\section{Introduction}

The theory of strong interactions, Quantum Chromodynamics (QCD), predicts a phase transition between hadronic gas and quark-gluon plasma (QGP) at high energy density. 
Lattice QCD predicts a critical temperature for such phase transition $T_c \approx 170$~MeV\cite{karsch}. An experimental test of the QCD prediction is being performed at the Relativistic Heavy Ion Collider, where the main focus of the physics  program   is to study the formation and characteristics of the QGP. 
The STAR experiment at RHIC has measured  a variety of hadron species ($\pi^{\pm}$, $\pi^0$, $K^{\pm}$, $K^0_s$, $K^*$, $\phi$, $p$, $\bar{p}$, $\Lambda$, $\bar{\Lambda}$,  $\Xi$, $\bar{\Xi}$, $\Omega+\bar{\Omega}$) in  Au+Au collisions at 200 GeV\cite{s1,s2,s3,PRL}.  
By  detailed investigation  of the identified particle spectra we study the  system properties at various stages of these collisions. 
  
\section{Data Analysis}

All measurements used for this work were performed by the STAR experiment. STAR main detector, time projection chamber\cite{starNIM},  is surrounded by a solenoidal magnet, which provided uniform magnetic field of 0.5T. Two zero-degree calorimeters were used for minimum bias triggering.
Minimum bias event sample was  divided  9  centrality classes based on measured charged particle multiplicity within pseudo-rapidity  $| \eta | < 0.5$; for rear particles some of the neighboring centrality bins were combined to improve statistics. 
A few different techniques were used for particle identification: correlation between the measured momentum and the specific ionization energy loss in the TPC gas was exploited to distinguish  pions, kaons and (anti)protons;  decay topology and invariant mass reconstruction were used to identify $K^0_s$,  $\Lambda$, $\bar{\Lambda}$, $\Xi$, $\bar{\Xi}$, and  $\Omega+\bar{\Omega}$; combinatoric invariant mass reconstruction technique was employed for $\phi$ and $K^*$ identification.
Corrections were applied to account for  tracking inefficiency, detector acceptance, hadronic interactions, and particle decays.  
 Pion spectra were also corrected for weak decay products; the rest of  spectra represent inclusive measurements. 

\section{Final state kinetic properties}
Particle transverse momentum distributions measured at RHIC show a  mass dependent hardening most pronounced in central collisions.
This hardening indicates the  presence of strong collective expansion (or flow) at RHIC.
Spectral shapes of particles  in a thermal, collectively expanding system, can be parametrized in the hydrodynamically motivated blast-wave model\cite{BW} by
kinetic freeze-out temperature $T_{kin}$, radial flow velocity $\beta$,  and a  flow velocity profile.
A single set of freeze-out parameters fits well all common particle spectra (pions, kaons and (anti)protons) at each centrality. Figure~1 shows the extracted kinetic freeze-out temperature and mean transverse flow velocity. A systematic decrease in $T_{kin}$ and increase in $\beta$ with centrality are observed, indicating a   more explosive, longer expanding system in central collisions.

\begin{figure}[ht]
\centerline{\epsfxsize=3.6in\epsfbox{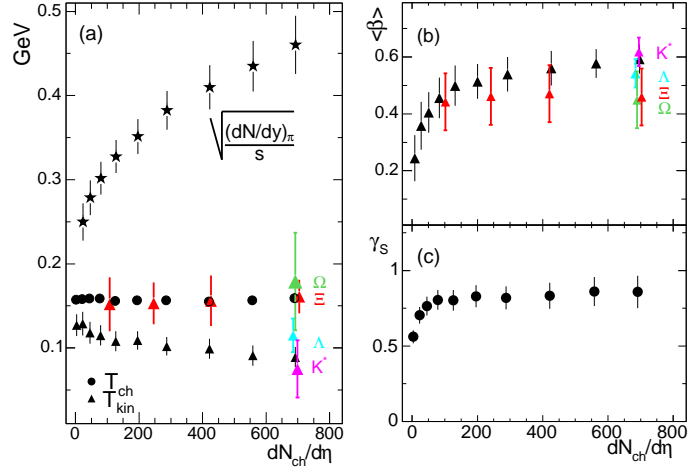}}   
\caption{ 
(a) $\sqrt{\frac{(\dndy)_{\pi}}{S}}$
(stars), $T_{ch}$ 
(circles) and $T_{kin}$ 
(triangles); 
 (b) $\langle \beta \rangle$ and
(c) $\gamma_s$
as a function of the charged  hadron multiplicity. 
Black symbols represent results extracted from $\pi$, $K$, $p$ measurements and shown with systematic errors.
Preliminary results for rare particles are shown in colored symbols (as marked on the plot),  error bars are statistical only.}
\end{figure}

The same set of freeze-out parameters,  describing the  common particle spectra, fails to reproduce other particle species measured by STAR;  spectral shapes of rare particles, fitted by the blast-wave model, reveal a different set of freeze-out parameters.  Freeze-out parameters extracted from single spectra fits of rare particles (shown in colored symbols in Figure~1) indicate sequential kinetic freeze-out of particle species: $\Omega$, $\Xi$,$\phi$ $\rightarrow$ $\Lambda$, $\pi$,$K$,$p$,$K^*$, likely happening  due to their smaller interaction cross-section with the bulk of the collision zone\cite{MSB}.

\section{Particle ratios and chemical freeze-out}
We use the blast-wave model results to extrapolate our spectra and obtain integrated $\dndy$ values.
We construct particle ratios to study  the chemical properties of the collision.
We fit our measured ratios with chemical equilibrium model\cite{new2,Nu}.  Extracted set of freeze-out parameters\cite{PRL} shows that baryon chemical potential  is independent of collision centrality within errors; strangeness chemical potential is consistent with 0. The obtained strangeness suppression factor $\gamma_{s}$ increases quickly from peripheral to mid-central collisions approaching unity for most central collisions (Figure~1c). This indicates  equilibration of strangeness  in central heavy-ion collisions at RHIC.
No centrality dependence is observed for chemical  freeze-out temperature $T_{ch}$ (Figure~1a) and its' value  is  close to the predicted  phase transition critical temperature. We also note that multi-strange baryons kinetic freeze-out temperature coincides with chemical freeze-out temperature for all centrality bins, suggesting that rare particles  decouple  from the system possibly at/right after chemical freeze-out. This allows one to estimate radial flow velocity at chemical freeze-out to be on the order of $0.45\pm0.10$~$c$. 

To characterize initial condition we use the following variable: $\frac{(dN/dy)_{\pi}}{S}$, where $S$ is an estimate of the transverse overlap area  based on the number of participants\cite{pi_star}.  This variable is related to the Bjorken estimate of energy density in the initial collision stage\cite{Bjorken}. It is also a relevant quantity in the parton saturation picture\cite{zxu2}, which suggests that in high energy collisions the initial gluon density is saturated up to a momentum scale that is proportional to $\sqrt{ \frac{(dN/dy)_{\pi}}{S}}$. 
Strong increase in $\sqrt{\frac{(dN/dy)_{\pi}}{S}}$ with collision centrality (shown in Figure~1a)  indicates higher energy densities achieved in   central collisions with respect to peripheral collisions. Contrasting system thermal properties at the three different stages (Figure~1a) we conclude that Au+Au collisions of different initial conditions at RHIC  always evolve toward the same chemical freeze-out temperature; chemical freeze-out is followed by further expansion and cooling,  with larger final radial flow in central collisions reflecting higher  energy densities at initial stage.

\begin{figure}[ht] 
\centerline{\epsfxsize=2.5in\epsfbox[0 50 567 470]{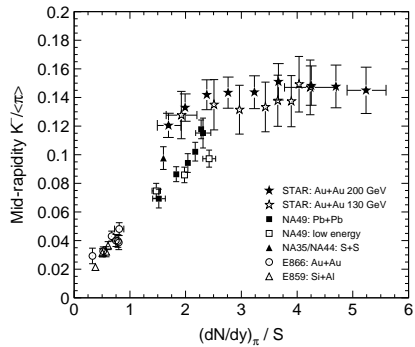}}    
\caption{  
Mid-rapidity $K^{-} / \pi$ ratio as function  of $\frac{(\dndy)_{\pi}}{S}$.  
Systematic errors are shown for STAR data, and statistical errors for other data. 
} 
\end{figure}
We also use $\frac{(\dndy)_{\pi}}{S}$ to compare strangeness  production at different energies. Strangeness enhancement has been long suggested as one of the possible signatures of a phase transition. Kaons carry more than 80\% of the strange quarks produced in a collision hence the $K/\pi$ ratio can be used to study strangeness production.
$K^-/\pi$ ratio is less affected by the degree of baryon stopping than $K^+/\pi$, and is therefore used in this study. The observed centrality independence of  $\kpi$ (also reflected by $\gamma_s$ in Figure~1c) is in  contrast to low energy data at SPS\cite{NA49} and AGS\cite{E802}.
We show results for different energies plotted as a function of $\frac{(\dndy)_{\pi}}{S}$ in Figure~2.
The trend that saturates at RHIC energies may be interpreted as  that strangeness production at low energies depends on how the collision was initially  prepared, but  not at RHIC energies.

\section{Conclusions}
Variety of hadron species measured by STAR in 200 GeV Au+Au collisions were studied within the framework of chemical and local kinetic equilibrium models to  investigate final  hadronic state properties of heavy ion collisions at RHIC. 
The results show that $T_{ch}$ is insensitive to centrality despite of changing initial conditions. The estimated value of $T_{ch}$
is close to the QCD predicted phase transition temperature. Kinetic freeze-out temperature of rare particles seems to coincide  with chemical freeze-out temperature for all centralities. 
Detailed study  of hadron spectral shape within blast-wave model indicates sequential kinetic freeze-out of particle species. Final kinetic freeze-out properties are correlated with centrality; 
higher flow velocity and lower temperature in central collisions seems to be caused by longer expansion due to higher initial energy density.

\vfill\eject
\end{document}